\documentclass[letterpaper, 10 pt, conference]{IEEEtran}
\IEEEoverridecommandlockouts
\usepackage{cite}
\usepackage{amsmath,amssymb,amsfonts}
\usepackage{algorithmic}
\usepackage{graphicx}
\usepackage{textcomp}
\usepackage{xcolor}
\usepackage{float}
\usepackage{multirow}

\usepackage{dirtytalk}
\def\BibTeX{{\rm B\kern-.05em{\sc i\kern-.025em b}\kern-.08em
    T\kern-.1667em\lower.7ex\hbox{E}\kern-.125emX}}

\begin{document}

\title{Spatial and Temporal Attention-based emotion estimation on HRI-AVC dataset\\
\thanks{* Authors contributed equally to this research article.

This material is based upon work supported by the National Science Foundation under Award No. DGE-2125362. Any opinions, findings, and conclusions or recommendations expressed in this material are those of the author(s) and do not necessarily reflect the views of the National Science Foundation.}
}

\author{\IEEEauthorblockN{Karthik Subramanian*}
\IEEEauthorblockA{\textit{Dept.  Elect. \& Microelectronic Eng.} \\
\textit{Rochester Institute of Technology}\\
Rochester, USA \\
kxs8997@rit.edu}
\and
\IEEEauthorblockN{Saurav Singh*}
\IEEEauthorblockA{\textit{Dept.  Elect. \& Microelectronic Eng.} \\
\textit{Rochester Institute of Technology}\\
Rochester, USA \\
ss3337@rit.edu}
\and
\IEEEauthorblockN{Justin Namba}
\IEEEauthorblockA{\textit{Dept. Computer Science} \\
\textit{Rochester Institute of Technology}\\
Rochester, USA \\
jrn1325@rit.edu}
\and
\IEEEauthorblockN{Jamison Heard}
\IEEEauthorblockA{\textit{Dept.  Elect. \& Microelectronic Eng.} \\
\textit{Rochester Institute of Technology}\\
Rochester, USA \\
jrheee@rit.edu}
\and
\IEEEauthorblockN{Christopher Kanan}
\IEEEauthorblockA{\textit{Dept. Computer Science} \\
\textit{University of Rochester}\\
Rochester, USA \\
ckanan@cs.rochester.edu}
\and
\IEEEauthorblockN{Ferat Sahin}
\IEEEauthorblockA{\textit{Dept.  Elect. \& Microelectronic Eng.} \\
\textit{Rochester Institute of Technology}\\
Rochester, USA \\
feseee@rit.edu}
}

\maketitle

\begin{abstract}
Many attempts have been made at estimating discrete emotions (calmness, anxiety, boredom, surprise, anger) and continuous emotional measures commonly used in psychology, namely `valence' (The pleasantness of the emotion being displayed) and `arousal' (The intensity of the emotion being displayed). Existing methods to estimate arousal and valence rely on learning from data sets, where an expert annotator labels every image frame. Access to an expert annotator is not always possible, and the annotation can also be tedious. Hence it is more practical to obtain self-reported arousal and valence values directly from the human in a real-time Human-Robot collaborative setting. Hence this paper provides an emotion data set (HRI-AVC) obtained while conducting a human-robot interaction (HRI) task. The self-reported pair of labels in this data set is associated with a set of image frames. This paper also proposes a spatial and temporal attention-based network to estimate arousal and valence from this set of image frames. The results show that an attention-based network can estimate valence and arousal on the HRI-AVC data set even when Arousal and Valence values are unavailable per frame.

\end{abstract}


\section{Introduction}
\label{sec_intro}

Research in facial emotion estimation using computer vision has increased in recent years. First attempts using Computer Vision to determine facial emotions relied on predicting discrete emotional classes such as calmness, anxiety, boredom, surprise, and anger \cite{canal2022survey}. While these methods performed well in determining discrete emotional categories, they failed to capture the entire  emotional spectrum experienced by humans. Thus, modern psychology uses dimensional models, such as the circumplex model, to determine the full range of human emotion \cite{barrett1998discrete,posner2005circumplex}. The circumplex model contains two axes, namely arousal and valence. Valence explains the pleasantness of emotion, whereas arousal captures the emotion's intensity. 

The attempts at creating data sets for detecting emotions as either categorical classes or containing arousal and valence values are broadly categorized in two ways: \textit{controlled} or \textit{in the wild}. \textit{In the wild} data sets refer to data that represent more natural conditions. AffectNet is the most extensive collection of images in the wild. It contains about one million images obtained from three major search engines. Human experts annotate nearly half of those images with arousal and valence values \cite{mollahosseini2017affectnet}. Another such data set is the SEWA \cite{kossaifi2019sewa}, which contains a rich diversity in the demographics of the images collected. The AFEW-VA \cite{kossaifi2017afew} is another similar example of an image data set with accurately labeled arousal and valence values. It consists of six hundred challenging video clips where human experts label each frame of the clip. The current state-of-the-art deep-learning-based computer vision methods can accurately and continuously estimate arousal and valence in naturalistic or \textit{In the wild} environments \cite{wang2022systematic, kossaifi2017afew, kossaifi2019sewa, toisoul2021estimation}.

This paper provides an emotion data set (HRI-AVC) obtained while conducting a human-robot interaction (HRI) task. The task involved a human and a robot working together to complete a repetitive assembly task while the human's face was continuously recorded. At the end of every repetition, humans were required to self-report their arousal and valence. Hence, this data set's arousal and valence values are associated with a group of image frames captured during each repetition. The experiment and the data set are described in more detail in Section \ref{sec_datset}.

EmoNet \cite{toisoul2021estimation} is one of the state-of-the-art methods that can estimate eight discrete emotion classes (neutral, happy, sad, surprise, fear, disgust, anger, contempt) and continuous emotion values (arousal and valence). The EmoNet was trained on the AffectNet data set \cite{mollahosseini2017affectnet}. Unlike typical emotion estimation networks, EmoNet extracts the facial landmarks and uses the landmarks for face alignment using the sub-network called the face alignment network (FAN) \cite{bulat2017far}. It further uses a combination of 2D convolution and dense layers to estimate the emotion classes, valence, and arousal. The HRI-AVC data set was tested using EmoNet. However, EmoNet does not perform well on the HRI-AVC dataset directly. To use EmoNet on the HRI-AVC data set, would require the availability of per-frame arousal and valence labels. The HRI-AVC data set has a pair of  valence-arousal values that represent the emotions of the human over a short time span which contains a set of image frames. The poor performance of EmoNet on the HRI-AVC dataset can be accounted to assuming that the same labels reported by the user are applied to all the frames in a set. To combat this problem this paper proposes a spatial and temporal attention-based network for estimating valence and arousal. The proposed network is built upon the existing spatial attention-based EmoNet architecture and aggregates embeddings obtained from multiple image frames to focus on temporal attention, essentially using EmoNet to provide features to learn on. The proposed method was trained and tested on the HRI-AVC data set and the results show that it can successfully estimate valence and arousal values with a sparsely labeled data set. The \textbf{key contributions} of this research are:
\begin{itemize}
    \item The creation of the HRI-AVC emotional data set for a collaborative human-robot assembly task.
    \item The use of a novel spatial and temporal attention-based emotion estimation network that can learn from samples with one valence-arousal value associated with a set of images.
\end{itemize}

The rest of the paper is organized as follows: Section \ref{sec_experimental_design} lays down the details of the experimental design used to collect the HRI-AVC dataset. Section \ref{sec_methodology} proposes an attention-based emotional network architecture. Section \ref{sec_results_discussion} presents the experimental results and discusses the findings, and Section \ref{sec_conclusion} concludes this paper.

\section{Experimental Design}
\label{sec_experimental_design}

A human-robot collaboration task experiment was conducted to generate a data set where each valence-arousal value pair has multiple associated images. The collaborative task was based on the investigation conducted by C. Savur et al.\cite{sahin2022evaluation} with the addition of a camera to record the participant's facial features. The experiment aimed to investigate the effects of human-robot interaction on human perceived safety, including emotions and comfort. Robot's behavior was manipulated to induce different ranges of human emotions by manipulating the \textit{velocity}, \textit{trajectory}, and \textit{sensitivity} of the robot. Participants perceived valence, arousal, and comfort were recorded after every interaction with the robot.

\begin{figure}[htbp]
\centerline{\includegraphics[width = 0.85\linewidth]{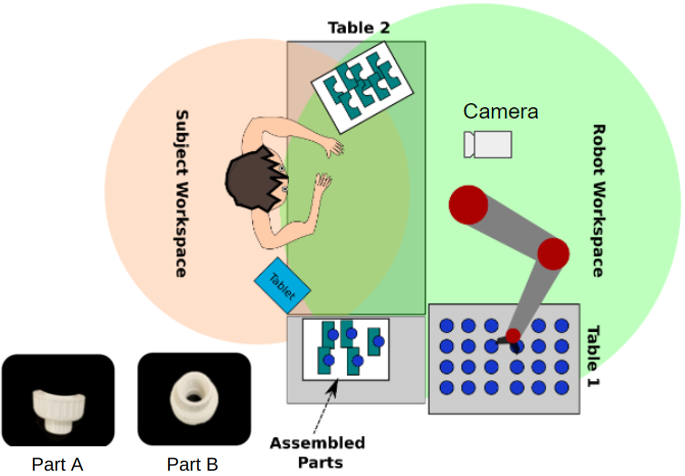}}
\caption{Experimental design used to collect the HRI-AVC dataset. \cite{sahin2022evaluation}}
\label{fig_exp_design}
\end{figure}

\begin{figure}[htbp]
\centerline{\includegraphics[width = 0.6\linewidth]{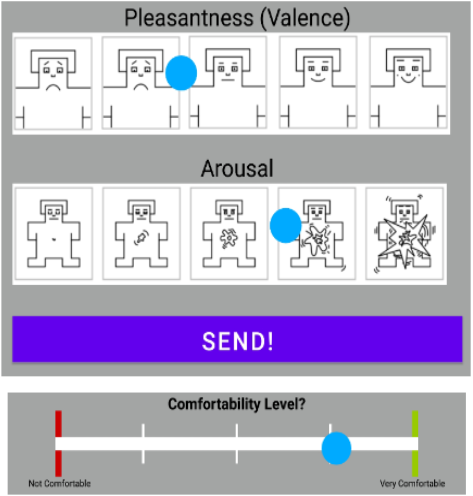}}
\caption{Manikin based GUI to capture Valence-Arousal subjective responses and perceived Comfort \cite{sahin2022evaluation}.}
\label{fig_manikin}
\end{figure}

The experimental design consisted of a human and a robot (Sawyer by Rethink Robotics) working on a collaborative task, see Figure \ref{fig_exp_design}. Each iteration of the experiment had a sequence of events, starting with the robot picking up an assembly part (A) from \textit{Table 1} and presenting the part to the participant seated in front of the robot. The participant picks up an assembly part (B) from \textit{Table 2} and assembles it to part (A) in the robot's grip. The robot then moves and drops the completed piece in the \textit{Assembled Parts} bin. Finally, the participant reports their perceived valence, arousal, and comfort for that iteration using the manikin-based GUI \cite{bradley1994measuring} on a tablet, Figure \ref{fig_manikin}. Throughout the iteration, a camera recorded the scene of interaction between the participant and the robot, capturing the facial features of the participant.

\begin{figure}[htbp]
\centerline{\includegraphics[width = \linewidth]{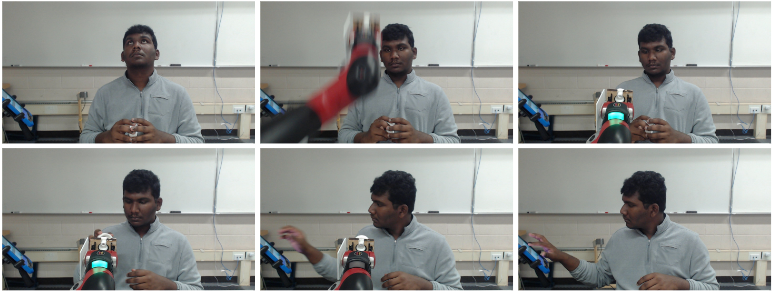}}
\caption{Example of image frames for an experiment iteration.}
\label{fig_example_image}
\end{figure}

The robot's velocity, trajectory, and sensitivity were manipulated to induce different levels of comfort and emotions in the participant. The velocity of the robot's end-effector can have two levels: Normal (N) and Fast (F). The robot's end-effector can take two possible trajectories: Normal (N) and Extreme (E). The robot's hand-off sensitivity can have two levels: Normal (N) and Sensitive (S). Hence, there are eight different combinations of the robot behavior parameters. Each participant goes through 10 trials where the first eight trials are the eight combinations of the robot behavior parameters in a fixed order, and the last two trials randomly use any of the eight combinations of robot behavior parameters. In each trial, the robot can provide up to 24 parts to assemble (iterations). After each iteration, the participant reports their subjective measures, labeling each iteration's recording with valence-arousal-comfort values. Figure \ref{fig_example_image} shows example image frames of a participant going through an iteration of a trial.


\subsection{HRI-AVC Dataset}
\label{sec_datset}
\begin{figure}[htbp]
\centerline{\includegraphics[width = 0.8\linewidth]{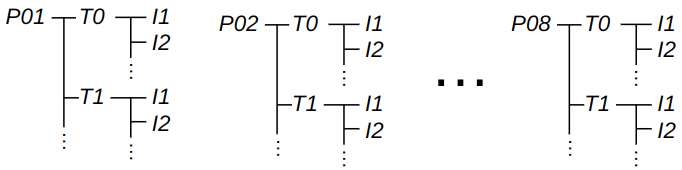}}
\caption{HRI-AVC dataset file structure.}
\label{fig_dataset_structure}
\end{figure}

The video and subjective measures data were stored in rosbag files. A rosbag is a file format used to store Robotic Operating System (ROS) messages. The video frames were extracted from when the robot picked up the assembly part (A) to when it dropped the assembled object, capturing all human responses during a random trajectory. These frames contain the information that explains valence-arousal-comfort values. Some video frames were extracted and stored in folders as images for each iteration. Each image folder includes the participant, trial, and iteration IDs. The folder name is structured as follows: \textbf{Person ID\_Trial ID\_Iteration ID}, see Figure \ref{fig_dataset_structure}. Participant ID represents the participant's ID and ranges from P01 to P12. Trial ID represents the number of times a participant completed 24 iterations and ranges from T00 to T09, where the T00 was the warm-up session. Iteration ID represents an iteration of the robot and the human collaboratively assembling an object and the human reporting their subjective responses. The total number of iterations for each subject should be 24; however, in some cases, the participants forgot to enter their subjective responses, so those iterations were removed from the dataset. Therefore, the number of iterations is less than 24 for some participants. A reference file that pairs folder names with associated valence-arousal-comfort values was also generated where the valence-arousal-comfort values range between 0 to 2.0. This file has four header columns: folder name, arousal, valence, and comfort. Thus, a data set was created where each valence-arousal value pair have multiple images in a folder associated with them.

\begin{figure}[htbp]
\centerline{\includegraphics[width = 0.76\linewidth]{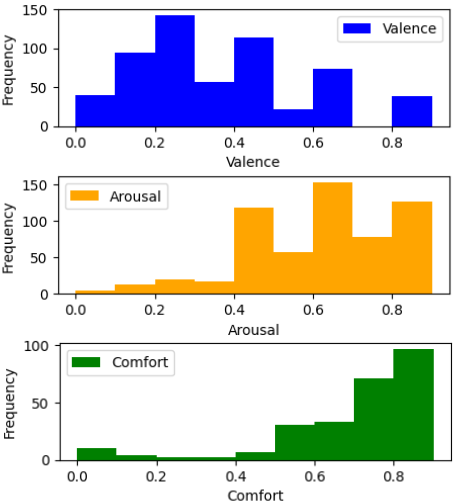}}
\caption{Histogram of the reported valence, arousal, and comfort in the HRI-AVC dataset.}
\label{fig_histogram}
\end{figure}

\begin{figure}[htbp]
\centerline{\includegraphics[width = 0.76\linewidth]{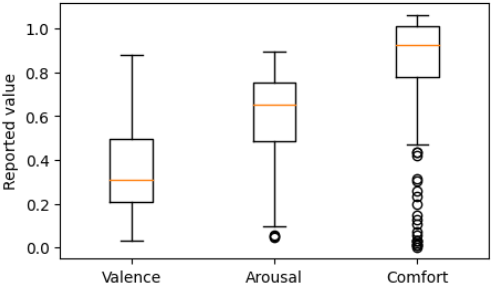}}
\caption{Boxplot of the reported valence, arousal, and comfort in the HRI-AVC dataset.}
\label{fig_boxplot}
\end{figure}

Six males and six females were recruited to work on the collaborative task with the robot. All the participants had minimal to no experience working with the robots. Data from 4 participants (P03, P05, P10, and P11) was removed from the data set due to inconsistencies in the data based on the post-experiment survey, which revealed their understanding of valence and arousal was incorrect. The valence-arousal-comfort values were rescaled between 0 to 1.0. Statistical analysis shows that the reported valence ranges from 0.031 to 0.880 with a mean of 0.253, and a standard deviation of 0.200. The reported arousal ranges from 0.045 to 0.891 with a mean of 0.622, and std of 0.171. Figure \ref{fig_histogram} and Figure \ref{fig_boxplot} shows the histogram and boxplot of the reported valence, arousal, and comfort in the HRI-AVC dataset.

\section{Methodology}
\label{sec_methodology}

\begin{figure*}[!htbp]
\centerline{\includegraphics[width = 0.9\textwidth]{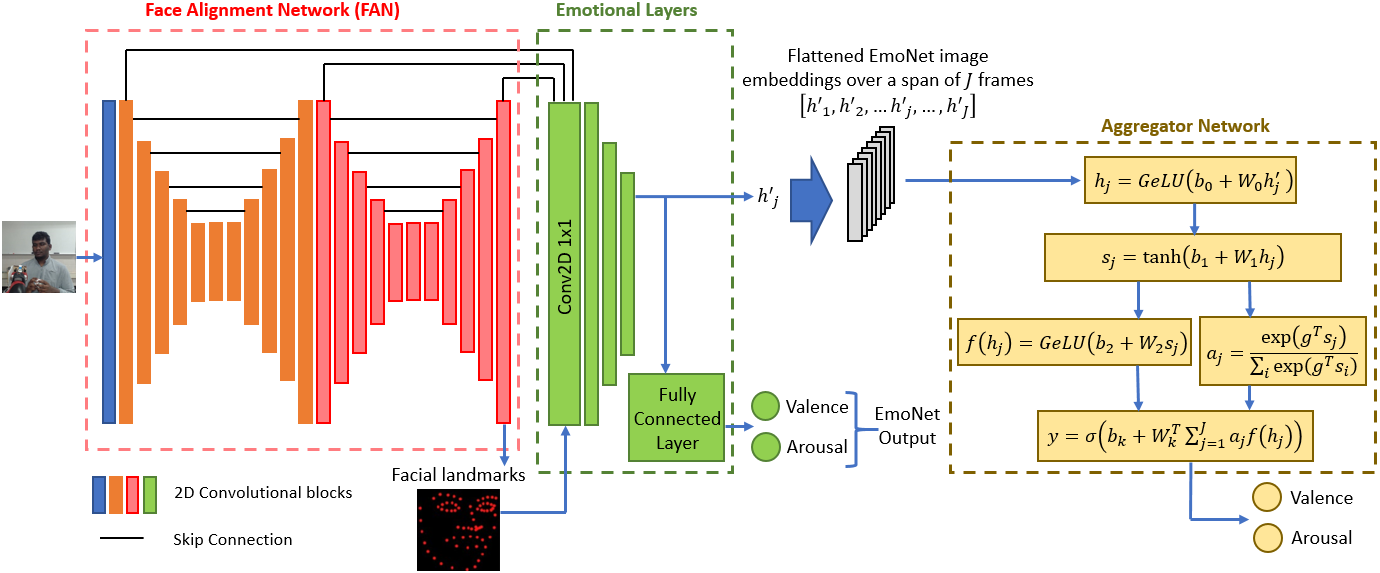}}
\caption{Attention-based Emotion Network architecture.}
\label{fig_aggnet}
\end{figure*}

There may be multiple faces with multiple scaling and orientations in an image. Traditionally, discrete categorical human emotions and continuous valence and arousal values are computed using a multistage process. A face detector detects the face in an image by finding the facial landmarks, then by re-scaling and aligning the beginning to estimate discrete and continuous human emotions.
EmoNet \cite{toisoul2021estimation} predicts the facial landmarks directly from an image and uses them for face alignment and emotion estimation, all within a single deep neural network. A subnet of EmoNet called a Facial Alignment Network (FAN) \cite{bulat2017far} is used to extract the facial features as a heat map. The emotional layers then use this heat map to extract features from the image. The extracted features are finally fed to the decision layers to estimate human emotions. EmoNet is designed to predict emotions based on a single image. This work addresses the problem of having a single valence-arousal value associated with multiple images. 

\subsection{Pre-processing}
The data collection experiment involved a collaborative task where the robot presented the participants with an assembly part (A). The participant assembles the assembly part (B) on part (A) and reports their perceived pleasantness (valence) and intensity (arousal). Since the participants look to their side to report their subjective measures, the most face gets occluded, which can be a potential source of error. Thus, the frames with a participant head deviation of more than 50 degrees from the center of the camera were removed from the data set.

\subsection{Attention-based Emotion Network}
This work proposes an attention-based emotion network, which uses multiple image frames with a single valence-arousal value associated with estimating continuous valence and arousal values. EmoNet image embeddings of numerous frames are aggregated using an aggregator network, thus adding a temporal attention mechanism to the existing spatial attention within the EmoNet. Figure \ref{fig_aggnet} shows the Attention-based Emotion Network architecture.

The attention-based emotion network uses the face alignment network and emotional layers from the EmoNet to generate flattened image embeddings represented by $h'_j$. EmoNet feeds the flattened image embeddings to fully connected layers to estimate valence and arousal for a single frame. However, the proposed attention-based emotion network computes the flattened image embeddings from a set of frames, the size of which is denoted by $J$, i.e., $J$ is the collective set of embeddings from image frames, $[h'_1,h'_2,\dots,h'_j,\dots,h'_J]$. These embeddings are passed through a fully connected layer to compute a linear projection from the CNN embeddings ($h_j$) and reduce the dimensional of the embeddings.

Next, attention mechanism \cite{vaswani2017attention} is used to aggregate the embeddings from $J$ frames. The embeddings are transformed using a fully connected layer with a tanh activation function:
\begin{equation}\label{eq_sj}
    s_j = \tanh(b_1 + W_1 h_j).
\end{equation}
$b_1$ and $W_1$ are learnable parameters. The attention values are calculated by applying a softmax on the score which is represented by ($g^Ts_j$), where $g$ is a trainable vector parameter.
\begin{equation}\label{eq_aj}
    a_j = \frac{\exp(g^Ts_j)}{\sum_i \exp(g^Ts_i)}
\end{equation}
$s_j$ is transformed into $f(h_j)$ using a fully connected layer with a GeLU activation function.
\begin{equation}\label{eq_fhj}
    f(h_j) = GeLU(b_2 + W_2 s_j)
\end{equation}
Here $b_2$ and $W_2$ are learnable parameters.
Finally, the $J$ frames are aggregated into a context vector by computing a dot product of the $f(h_j)$, and the attention values $a_j$ ($\sum_{j=1}^J a_j f(h_j)$). The context vector is fed to another fully connected layer with a sigmoid activation function to estimate the valence and arousal values.

\begin{equation}\label{eq_y}
    y = \sigma \left( b_k + W_k^T \sum_{j=1}^J a_j f(h_j) \right)
\end{equation}
Here, $b_k$ and $W_k^t$ are learnable parameters. The aggregator network is trained for 50 epochs for creating individual models and 100 epochs for creating universal models. Optimizing the network is done with \textit{AdamW} optimizer.

\section{Results \& Discussion}
\label{sec_results_discussion}

The performance of the proposed network is evaluated using the Root Mean Squared Error (RMSE), Pearson Correlation Coefficients (PCC), and Concordance Correlation Coefficients (CCC) as the performance metrics.

Root Mean Squared Error evaluates how close the predicted values are to the ground truth values.
\begin{equation}
    RMSE(Y,\hat{Y}) = \sqrt{\mathbb{E}(Y - \hat{Y})^2)}
\end{equation}
Pearson Correlation Coefficients (PCC) evaluates the correlation between the predicted and the ground truth values.

\begin{equation}
    PCC(Y,\hat{Y}) = \frac{\mathbb{E}(Y - \mu_Y)(\hat{Y} - mu_{\hat{Y}})}{\sigma_Y \sigma_{\hat{Y}}}
\end{equation}
Concordance Correlation Coefficients (CCC) evaluate the agreeability between the predicted and the ground truth values. Unlike PCC, CCC considers both the correlation and the bias between the two values.

\begin{equation}
    CCC(Y,\hat{Y}) = \frac{2 \sigma_Y \sigma_{\hat{Y}}PCC(Y,\hat{Y})}{\sigma_Y^2 + \sigma_{\hat{Y}}^2 + (\mu_Y - \mu_{\hat{Y}})^2}
\end{equation}

Existing methods for detecting arousal and valence from images like EmoNet require learning from data sets containing labels for every frame. The HRI-AVC data set contain labels for a set of frames. If each image in a set of frames for each iteration is assigned the participant-reported label, existing methods fail to estimate the target outputs, as shown in Table \ref{table_emonet}. The results show that the estimated valence and arousal values have no correlation with the ground truth values and have high RMSE with HRI-AVC dataset. This suggests that there is a need to develop methods which predicts human emotional states over a window of time to be applied to more practical applications such as human robot collaboration.


\begin{table}[b]
\caption{EmoNet performance on different datasets \cite{toisoul2021estimation}}
\label{table_emonet}
\centering
\begin{tabular}{|c|ccc|ccc|}
\hline
\multirow{2}{*}{Dataset}                                       & \multicolumn{3}{c|}{Valence}                                                      & \multicolumn{3}{c|}{Arousal}                                                        \\ \cline{2-7} 
                                                               & \multicolumn{1}{c|}{CCC}  & \multicolumn{1}{c|}{PCC}  & RMSE                      & \multicolumn{1}{c|}{CCC}   & \multicolumn{1}{c|}{PCC}   & RMSE                      \\ \hline
\begin{tabular}[c]{@{}c@{}}AffectNet\\ (original)\end{tabular} & \multicolumn{1}{c|}{0.73} & \multicolumn{1}{c|}{0.73} & 0.33                      & \multicolumn{1}{c|}{0.65}  & \multicolumn{1}{c|}{0.65}  & 0.30                      \\ \hline
\begin{tabular}[c]{@{}c@{}}AffectNet\\ (clean)\end{tabular}    & \multicolumn{1}{c|}{0.82} & \multicolumn{1}{c|}{0.82} & 0.29                      & \multicolumn{1}{c|}{0.75}  & \multicolumn{1}{c|}{0.75}  & 0.27                      \\ \hline
\multicolumn{1}{|c|}{SEWA}                                     & \multicolumn{1}{c|}{0.65} & \multicolumn{1}{l|}{0.66} & \multicolumn{1}{c|}{0.32} & \multicolumn{1}{c|}{0.61}  & \multicolumn{1}{c|}{0.61}  & \multicolumn{1}{c|}{0.35} \\ \hline
AFEW-VA                                                        & \multicolumn{1}{c|}{0.69} & \multicolumn{1}{c|}{0.70} & 0.23                      & \multicolumn{1}{c|}{0.66}  & \multicolumn{1}{c|}{0.67}  & 0.22                      \\ \hline
\textbf{HRI-AVC*}                                                        & \multicolumn{1}{c|}{0.02} & \multicolumn{1}{c|}{0.03} & 0.43                      & \multicolumn{1}{c|}{-0.03} & \multicolumn{1}{c|}{-0.10} & 0.55                      \\ \hline
\end{tabular}
\end{table}


Individual models are created for each participant by training on two-thirds of the participant data obtained from the data collection process described in Section \ref{sec_experimental_design}. The proposed method is trained on the train set for a short period, and the obtained model is evaluated on the withheld test set. Table \ref{table_individual} shows the obtained results. Most learned individual models can understand the valence target output better than the arousal output; this is expected as there is some evidence of Humans being not very good at reporting arousal \cite{grimm2007robust,russell2003facial}.
Overall, it can be seen that this method shows that, at an individual level, it is possible to learn the emotional targets of arousal and valence. Some individual models are better than others. Of the eight participants, three learned models, namely models trained on Participants 1, 4, and 8, showed better than average valence learning. Participants 2, 6, 9, and 12 showed comparable to-average performance on their respective test sets. Only the model trained on participant seven's data had a very poor concordance correlation between the ground truth and predictions. All models show the Pearson Correlation coefficient, suggesting that it is learning the valence target output correctly. The performance for correctly evaluating the output target of arousal shows that four of the eight models show better than average performance. In contrast to learning valence, the remaining models offer weak to no Concordance correlation. Similar trends are observed for arousal. RMSE values for both target outputs are reasonably low.

Multi-fold cross-validation is performed on the entire data set to obtain a good-performing model representing all participants.
The best model was obtained when the training set included participants 1, 4, 6, and 8. Table \ref{table_universal} shows the results of evaluating this universal model on the data of participants not used in training. The test data includes participants 2, 7, 9 and 12. Each participant's valence prediction using both correlation metrics is comparable to the average performance. In contrast, arousal performs worse in three participants in the training data set. This is in agreement with the results of the individual models. Arousal metrics for participants 2, 7, and 9 were also poor using individual models. It can be noted that Participant 12 improved using the universal model compared to the individual model for the arousal metric, and the performance for the valence metric also suggests that the universal model has learned to make relevant predictions. 

Finally, the universal model is fine-tuned with data from each participant to create individual models for every participant in the test set. The performance of measuring valence improved  on all participant's over all metrics, as seen in Table \ref{table_fine_tuned}. Interestingly, Participant 7, whose individual model could not predict valence, could learn to predict the target output of valence better than previously seen on both the individual and universal models. Similar patterns of being unable to understand the target output of arousal are seen in participants 2, 7, and 9.

\begin{table*}[t]
\caption{Individual Models}
\label{table_individual}
\centering
\begin{tabular}{|c|c|c|c|c|c|c|}
\hline
\textbf{Participant \_ID} & \textbf{Valence CCC} & \textbf{Valence PCC} & \textbf{Valence RMSE} & \textbf{Arousal CCC} & \textbf{Arousal PCC} & \textbf{Arousal RMSE} \\ \hline
1                         & 0.415                & 0.485                & 0.219                 & 0.351                & 0.424                & 0.220                  \\ \hline
2                         & 0.187                & 0.45                 & 0.199                 & -0.055               & -0.113               & 0.238                 \\ \hline
4                         & 0.497                & 0.565                & 0.080                  & 0.704                & 0.791                & 0.082                 \\ \hline
6                         & 0.204                & 0.317                & 0.240                  & 0.127                & 0.164                & 0.128                 \\ \hline
7                         & 0.045                & 0.285                & 0.135                 & 0.033                & 0.286                & 0.070                  \\ \hline
8                         & 0.583                & 0.701                & 0.102                 & 0.308                & 0.659                & 0.125                 \\ \hline
9                         & 0.144                & 0.209                & 0.073                 & 0.048                & 0.081                & 0.054                 \\ \hline
12                        & 0.201                & 0.337                & 0.113                 & 0.365                & 0.676                & 0.132                 \\ \hline\hline

Mean & 0.285 & 0.418& 0.145& 0.235& 0.371& 0.131 \\ \hline
Std & 0.189 & 0.162& 0.065& 0.247& 0.321& 0.067
\\ \hline
\end{tabular}
\end{table*}

\begin{table*}[t]
\caption{Testing with hold out participants on the best Universal Model}
\label{table_universal}
\centering
\begin{tabular}{|c|c|c|c|c|c|c|}
\hline
\textbf{Participant ID} & \textbf{Valence CCC} & \textbf{Valence PCC} & \textbf{Valence RMSE} & \textbf{Arousal CCC} & \textbf{Arousal PCC} & \textbf{Arousal RMSE} \\ \hline
2                       & 0.1364               & 0.363                 & 0.254                  & -0.013                & -0.003               & 0.338                \\ \hline
7                       & 0.120                & 0.158               & 0.148               & 0.008               & 0.001               & 0.121                 \\ \hline
9                       & 0.118                & 0.235              & 0.088               & 0.049                & 0.009                & 0.080                  \\ \hline
12                      & 0.126                & 0.297                & 0.174                & 0.574               & 0.635                 & 0.101                 \\ \hline\hline

Mean & 0.125 & 0.263& 0.166& 0.154& 0.176& 0.160 \\ \hline
Std & 0.008 & 0.008& 0.006& 0.280& 0.311& 0.119
\\ \hline
\end{tabular}
\end{table*}

\begin{table*}[t]
\caption{Fine Tuned on Universal Models}
\label{table_fine_tuned}
\centering
\begin{tabular}{|c|c|c|c|c|c|c|}
\hline
\textbf{Participant ID} & \textbf{Valence CCC} & \textbf{Valence PCC} & \textbf{Valence RMSE} & \textbf{Arousal CCC} & \textbf{Arousal PCC} & \textbf{Arousal RMSE} \\ \hline
2                       & 0.226                & 0.350                 & 0.150                  & -0.03                & -0.014               & 0.116                 \\ \hline
7                       & 0.239                & 0.377                & 0.126                 & -0.019               & -0.031               & 0.122                 \\ \hline
9                       & 0.309                & 0.466                & 0.070                  & 0.003                & 0.006                & 0.070                  \\ \hline
12                      & 0.354                & 0.652                & 0.090                  & 0.670                 & 0.680                 & 0.109                 \\ \hline\hline

Mean & 0.282 & 0.461& 0.109& 0.156& 0.128& 0.104 \\ \hline
Std & 0.050 & 0.118& 0.030& 0.296& 0.322& 0.020
\\ \hline
\end{tabular}
\end{table*}

\section{Conclusions \& Future Work}
\label{sec_conclusion}
This paper shows that learning arousal and valence values over a small attention span is possible. It makes available a data set containing eight participants who self-reported their arousal and valence while working with the robot. This data set varies from other Arousal Valence data sets because the arousal and valence values are self-reported and not annotated by a third party. However, the labels of the target output are applied to a set of frames. This meant that conventional methods requiring arousal and valence values for every frame would not work. It was verified that out-of-the-box networks like EmoNet yielded inferior results on this data. To overcome this, a new aggregator network was introduced that utilized the embeddings from EmoNet and learned to predict arousal and valence over a small attention span. The results show that this method learns valence better than arousal on the data available, suggesting that learning arousal may require more data.

We intend to collect more data over time to increase the size of the data set to include at least forty participants to make it more diverse. Experiment with different architectures of the aggregator network to see if performance can be improved. Implement a real-time predictor that can be used in Human-Robot Collaboration applications. 

\bibliographystyle{IEEEtran}
\bibliography{bibliography}

\end{document}